\documentclass[prl,twocolumn,showpacs,preprintnumbers,amsmath,amssymb]{revtex4}

\usepackage{epsfig}

\begin{document}


\title{
Current-spin coupling for ferromagnetic domain walls in fine wires } 

\author{S. E. Barnes and S. Maekawa }
\affiliation{Institute for Materials Research, Tohoku University, 
Sendai 980-8577, Japan}

\date{\today} 
\begin{abstract}
{
The coupling between a current and a domain wall is examined.  In the presence of a finite current and the absence of a potential which breaks the translational symmetry, there is a perfect transfer of angular momentum from the conduction electrons to the wall. As a result, the ground state is in uniform motion. This remains the case when relaxation is accounted for. This is described by, appropriately modified, Landau-Lifshitz-Gilbert equations.
 }
\end{abstract}


\maketitle

Spintronic devices have great technological promise but represent a challenging problem at both an applied and fundamental level. It has been  shown theoretically\cite{1,2} that the direction of a magnetic domain might be switched  using currents alone. Devices designed to use this principle often consist of multi-layers of magnetic and non-magnetic conductors. The advantages of similar devices based upon the  current induced displacement of a {\it domain wall\/} are simplicity and the fact that the switching current is much smaller\cite{3,4,5,6}. Experimentally the current induced displacement of a domain wall has been clearly demonstrated and in recent experiments\cite{5,6} the velocity of the  wall was measured.  

The current induced motion of a magnetic domain  involves the transfer of angular momentum from the conduction electrons. The early theory\cite{1,2} and most of the subsequent work\cite{7} is based upon some type of assumption about this torque transfer process and there is no real consensus on how this should appear in the (Landau-Lifshitz-Gilbert) equations of motion\cite{7}. The purpose of this Letter is to develop a {\it complete\/} first principles theory of this process for a {\it domain  wall}, this based upon a specific model Hamiltonian and physically  justified approximations. This is intended to form the basis for  the more complicated problem of multi-layers.

Although similar conclusions are valid for the Stoner, and related, models, here attention will be focused upon the $s-d$-exchange model. The direction of the local moments, $\vec S_i$, which make up the domain wall are specified by the usual Euler angles $\theta_i$ and $\phi_i$. To make diagonal the interaction $+J\vec S_i \cdot \vec s_i$, at site $i$, the axis of quantization of the conduction electrons, $\vec s_i$, is rotated  along this same direction. If $ \psi(\vec r_i, t) $ is the conduction electron spinor field then this amounts to making a SU(2) gauge transformation, $ \psi(\vec r_i, t) \to r(\theta_i, \phi_i) \psi(\vec r_i, t)$, where $r(\theta_i, \phi_i) \equiv e^{i\phi_i s_{z}} e^{i\theta_i s_{y}} e^{i\phi_i s_{z}}  =  (\cos (\theta_i/2) + i \sin (\theta_i/2)\sin \phi_i \sigma_x - i \sin (\theta_i/2)\cos \phi_i \sigma_y)$ and where $\vec \sigma$ are the Pauli matrices.  If there was SU(2) gauge {\it invariance}, there would exist {\it three\/} gauge particles, the analogy of the $W^\pm$ and $Z$ which mediate the weak interaction. However here, without this invariance, this  transformation still introduces {\it three\/} gauge fields. The {\it longitudinal\/} such field has been exploited  in the  development of  theories of the Hall effect\cite{8}. It is Bazaliy et al.\cite{9}, who first  developed a  theory of the torque transfer to a magnetic  domain, for the half-metal limit,  based upon this same field. Their formalism generates a {\it transverse\/}  field in the Landau-Litshiftz equations through  a finite derivative $\partial {\cal V}_{}\{\theta_i, \phi_i\}/\partial \phi_i$ where $ {\cal V}_{}\{\theta_i, \phi_i\}$ is the expectation value of the energy as a function of the angles $ \{\theta_i, \phi_i\}$.
 {\it However},  a finite ground state expectation value  for such a derivative implies that 
 the solution is not stable and it follows that their  finite velocity solution must relax to one which is {\it stationary\/} with a finite common value of the $\phi_i =\phi_0$, this anticipating the conclusions of Tatara and Kohno\cite{10}. 
 
Here it will be shown that the torque transfer effects can be correctly accounted for in a simpler  U(1) theory in which the only generators are $S_{iy}$ for rotations about the axis  perpendicular to the wall plane. The local spins are usually co-planer, see below, and the  rotations $r(\theta_i) \equiv  e^{i\theta_i s_{y}}  =  (\cos (\theta_i/2)  - i \sin (\theta_i/2)\sigma_y)$ are all that is needed to diagonalize  $+J\vec S \cdot \vec s$. This simpler approach can {\it only\/} generate a  {\it transverse\/} gauge field, i.e., the one ignored in the earlier work\cite{9},  and which {\it alone\/} is found to be the origin of  the torque transfer process. The longitudinal field, along with $\partial {\cal V}_{}\{\theta_i, \phi_i\}/\partial \phi_i$, and the ground state $\phi_0$ are strictly null, and  the correctly {\it relaxed\/} ground state {\it does\/} have a finite velocity  in the absence of pinning.

The $s-d$-exchange  Hamiltonian is,
\begin{eqnarray}
&&{\cal H}
= - \sum_{<ij>\sigma \sigma^\prime}\left(c^\dagger_{i\sigma}t_{ij\sigma \sigma^\prime}c^{}_{j\sigma^\prime} + H.c.\right) - \mu \hat N
\nonumber \\
 &&- \sum_{i} (J\vec S_i\cdot \vec s_i +A^0 {S_{iz}}^2 - K^0_\perp {S_{iy}}^2)
 \nonumber \\
 &&
-  J^0_s \sum_{<ij>}
\vec S_i\cdot \vec S_j 
\label{un}
\end{eqnarray}
where $c^\dagger_{i\sigma}$  is the conduction electron creation operator for spin $\sigma$ and site $i$. The uniform hopping integral is $t_{ij\sigma \sigma^\prime} = t\delta_{\sigma \sigma^\prime}$ (the strange notation is for later use), and $\hat N$ is number operator. The exchange constants are  $J$ and $J^0_{s}$ for the coupling between the local moments and the conduction electrons and for the direct coupling between local moments. With the anisotropy constant $A^0>0$ the $z$ direction is an easy axis  while, with $K^0_\perp>0$, the $y$ axis is hard. These anisotropy constants include both intrinsic anisotropy plus the effects of a demagnetizing field. 

It is first necessary to describe the domain wall. The Holstein-Primakoff transformation\cite{11} $S_{iz}=S-b^\dagger_i b^{}_i$, $S^+_i = (2S- b^\dagger_i b^{}_i)^{1/2}b^{}_i \approx (2S)^{1/2}b^{}_i$ is used to quantize the spin degrees of freedom using the direction defined by $\theta_i$ and $\phi_i$ as the axis of quantization for the local spins. To within a constant, for the spin Hamiltonian in Eqn.~(\ref{un}),  ${\cal V}_{}\{\theta_i, \phi_i\}   = - AS^2 \cos^2 \theta_i + K_\perp S^2 \sin^2 \theta_i  \sin^2 \phi_i -  \sum_{<ij>}J_{s} \cos \theta_{ij} $ where $A = [(2S-1)/2S]A^0$,  $K_\perp = [(2S-1)/2S]K^0_\perp$, and  $\theta_{ij}$ is the angle between the spins at sites $i$ and $j$. The energy $K_\perp S^2 \sin^2 \theta_i  \sin^2 \phi_i $ favors a wall with $\phi_i =0$, whence assuming a slow variation in the angle $\theta_i$, i.e., that  $\theta_{ij} \approx \nabla \theta_i \cdot \vec r_{ij}$, where $\vec r_{ij}= \vec r_j - \vec r_i$ is the vector which joins near neighbor planes of the wall in the $z$ direction ($r_{ij}=a$), and again dropping a constant ${\cal V}_{}\{\theta_i\}   = +J_s S^2 ( \nabla \theta_i \cdot \vec r_{ij})^2 - AS^2 \cos^2 \theta_i$. The domain wall structure, which lies in the $x-z$-plane,  can be determined by minimizing this ${\cal V}_{}\{\theta_i\}$, and the standard result is $\theta(z_i) =2\cot^{-1} e^{-(z_i/w)}$, where $z_i$ is coordinate of site $i$. The wall width $ w= a(J_s / 2A)^{1/2}$. The local spins rotate so $\theta = 0$ becomes $\theta = \pi$ with increasing $z$.

In order to have precise expressions for the conjugate position and momentum of the wall, it is observed   that the later is the generator of displacements. The product of small $y$-axis rotations $R(\Delta z)=\prod_i \exp ( i (\theta_i - \theta_{j})(\Delta z/a)(S_{iy}/\hbar))\approx  (1+ (1/2)\sum_i (\nabla \theta_i \cdot \vec r_{ij})\Delta (z/a)(2S)^{1/2}(b^\dagger_i - b^{}_i))$, produces a wall translation by $\Delta z$, and  so the momentum, per atom in a plane of the wall,   $p_z = i (\hbar /2) (v_c/a^2{\cal A}) \sum_i (\nabla \theta_i \cdot \vec r_{ij})(2S)^{1/2} (b^\dagger_i - b^{}_i) $, where $v_c$ is the cell volume and ${\cal A}$ is the cross-sectional area of the wire. The Goldstone boson which restores the translational invariance of $\cal H$ can then be identified as,
$$
b^\dagger_w =({J / 8A})^{1/4}(v_c/ a{\cal A}) \sum_i (\nabla \theta_i \cdot \vec r_{ij})\, b^\dagger_i,
$$ 
where the constant of proportionality  reflects the requirement that $[b^{}_w, b^\dagger_w]=1$ and uses the explicit wall solution for $\theta(z)$. (The  modes which disperse from this Goldstone boson are excitations of the wall.) Thus $\hat p_z = i \hbar({2A / J})^{1/4} (2S)^{1/2}(1/a)(b^\dagger_w - b^{}_w)$ while the conjugate coordinate, such that $[\hat z_0, \hat p_z]=i\hbar$, is $\hat z_0 = a ({J / 2A})^{1/4} (S/2)^{-1/2} (b^\dagger_w + b^{}_w)$. 

Returning to the description of the conduction electrons, performing the above described SU(2) rotations, the conduction electron part of Eqn.~(\ref{un})  reduces to,
\begin{equation}
{\cal H}_e
= - \sum_{<ij>\sigma \sigma^\prime }\left(c^\dagger_{i\sigma}t_{ij\sigma \sigma^\prime}c^{}_{j\sigma^\prime } + H.c.\right) 
- JS \sum_{i} s_{iz} - \mu \hat N,
\label{troisbis}
\end{equation}
in which now $t_{ij} = t r^{-1}(\theta_i, \phi_i)r(\theta_j, \phi_j)$.
But since  $\phi_i=0$ this simplifies to the U(1) result:
\begin{equation}
t_{ij} = t (\cos (\theta_i/2)  + i \sin (\theta_i/2)\sigma_y)(\cos (\theta_j/2)  - i \sin (\theta_j/2)\sigma_y).
\end{equation}
which  is,
$
t_{ij}= t[\cos (\theta_i/2) \cos (\theta_j/2) + \sin (\theta_i/2) \sin (\theta_j/2) 
+  i (\sin (\theta_i/2)\cos (\theta_j/2)- \sin (\theta_j/2)\cos (\theta_i/2)\sigma_y]
$, using  ${\sigma_y}^2 =1$.
 The standard identities  then lead to $t_{ij} = t [\cos \nabla \theta_i \cdot \vec r_{ij} + i (\nabla \theta_i \cdot \vec r_{ij}) \sigma_y]$, and correct to second order in the gradient, this  can be written as,
\begin{equation}
t_{ij} =e^{i\int_{\vec r_i}^{\vec r_j}  \vec A \cdot d\vec r} t  \cos (\nabla \theta_i \cdot \vec r_{ij})  ; \ \ \   \vec A = (\nabla \theta_i \cdot \vec r_{ij})s_y \hat z
\label{quatre}
\end{equation}
In  order that this central result  be valid, the conduction electrons must follow the local spin magnetization as they pass through the wall and this requires that the adiabatic theorem be satisfied.  
When the transverse field represented by $\Delta t^\perp_{ij} = i t ( \nabla \theta_i \cdot \vec r_{ij} ) \sigma_y$ is ignored (the equivalent to $\vec A=0$), the eigenstates of ${\cal H}_e$ are also  eigenstates of the total angular momentum $\sum_i(\vec S_i + \vec s_i)$ evaluated in the {\it local\/} frame. Given that $J>0$, in the ground state, $\vec S_i$ and  $\vec s_i$ {\it are\/} then parallel. 

This approximation is manifestly valid in the half metal limit when $J \gg t$. The ground state is then a mixture of states in which all sites are either singly occupied by an electron or unoccupied. The singly occupied sites with the maximum angular momentum $S+(1/2)$ have the lowest energy while other states, and those with two electrons per site, have an energy which is higher by $\sim J$ and hence have negligible weight in the ground state. The Wigner-Eckart theorem then dictates that all the matrix elements of $\vec s_i$ are equal to those of $(\vec S_i/2S)$.  

However, since spatially the magnetization  rotates slowly, a much weaker inequality suffices. The adiabatic theorem simply demands that the transverse field  $\Delta t^\perp_{ij} = i t  (\nabla \theta_i \cdot \vec r_{ij} )\sigma_y$  be small compared to the longitudinal field $Jm_s$.  The wall rotates by $\pi$ over a distance $w$ so $\nabla \theta_i  \sim \pi/w$ and $\Delta t^\perp_{ij} \sim i \pi t (a/w)$, and required is,
\begin{equation}
\pi t (a/w) \ll Jm_s
\label{cinq}
\end{equation}
which since, e.g., for Permalloy $w/a \sim 10^3$ is typically well satisfied. 
The conduction electron magnetization comprises two components with, by definition, the (minority) majority  conduction electrons  (anti-) parallel to the axis of quantization, i.e.,  the direction of the local  spin. In the {\it local frame\/} the majority (minority) electrons have $\sigma_z = +1$ ($\sigma_z = -1$), so that it follows that when the adiabatic theorem is satisfied,
\begin{equation}
\vec s_i = \sigma_z (\vec S_i/2S),
\label{quatrebis}
\end{equation}
independent of the details of the electronic structure, etc. 

The non-interacting approximation is therefore to indeed ignore $\Delta t^\perp_{ij}$. If it was not for the cosine in $t_{ij} =t  \cos (\nabla \theta_i \cdot \vec r_{ij}) $ the problem is then identical to that without the wall. Using again $(\nabla \theta_i \cdot \vec r_{ij}) \sim \pi a/w$, the correction is $\sim t \pi^2 (a/w)^2$ which with $(a/w) \sim 10^{-3}$  might be safely ignored, i.e., at this level the electronic structure is unchanged by the wall and since the reflection probability is small, there is a negligible {\it pressure\/} exerted by the conduction electrons. However, in the spin sector, this correction {\it is\/} important. Evaluating the coefficient of $(\nabla \theta_i \cdot \vec r_{ij}) ^2$ in first order perturbation theory  leads to, the renormalization $J_s = J^0_s + (x^\prime t/2S^2)$, of the exchange coupling\cite{f2}. The effective concentration $x^\prime = \langle c^\dagger_i c^{}_j \rangle$.

At this lowest level of approximation the wall is stationary and  is put in motion only when the torque transfer $\Delta t^\perp_{ij}$-term is accounted for as a perturbation. Even with this term, the effective Hamiltonian, Eqn.~(\ref{troisbis}), is evidently of single particle nature. 
Consider first the single particle description of the majority spin electrons. In order to account for the torque transfer term proportional to $\Delta t^\perp_{ij}$, use is made of Eqn.~(\ref{quatrebis}). To this end, it is noted, for  majority electrons  $\sum_{\sigma \sigma^\prime}c^\dagger_{i\sigma}\sigma_{y\sigma \sigma^\prime}c^{}_{j\sigma^\prime} =i(s^-_{i}c^\dagger_{i\uparrow} c^{}_{j\uparrow} - c^\dagger_{i\uparrow} c^{}_{j\uparrow}s^+_{j})$. Then by virtue of Eqn.~(\ref{quatrebis}), e.g., $s^+_{i} \approx  (2S)^{-1/2} b^{}_i$, and $\sum_{\sigma \sigma^\prime}c^\dagger_{i\sigma}\sigma_{y\sigma \sigma^\prime}c^{}_{j\sigma^\prime} = i(2S)^{-1/2} c^\dagger_{i\uparrow} c^{}_{j\uparrow}(b^\dagger_{i}-b^{}_{j})$. Combining this with the similar result for the minority electrons, the $\Delta t^\perp_{ij}$-term becomes,
\begin{equation}
{\cal H}_{\tau} = 
-\frac{t}{2(2S)^{1/2}} \sum_{ij\sigma} \sigma (\nabla \theta_i \cdot \vec r_{ij})
c^\dagger_{i\sigma} c^{}_{j\sigma} (b^{}_i - b^{\dagger}_{j}) + H.c.,
\label{deuxbis}
\end{equation}
which couples the spin current to the magnons and reflects the {\it entire\/} torque transfer process. The necessary correction is obtained, directly, by taking the expectation value with respect to the conduction electrons, whence Eqn.~(\ref{deuxbis})  reduces to
\begin{equation}
{\cal H}_{\tau} = 
 -i \frac{ \hbar j_s a^2 }{ 2 e S}
  \sum_i (\nabla \theta_i \cdot \vec r_{ij})  (2S)^{1/2}  (b^{}_i - b^{\dagger}_{i}),
\label{troisbisbis}
\end{equation}
where $j_s$ is the {\it spin\/} current. Here  $b^{\dagger}_{i+1}$ is replaced by $b^{\dagger}_{i}$, an approximation  valid  in the continuum limit  $w\gg a$.  The quantity $(2S)^{1/2}  (b^{}_i - b^{\dagger}_{i}) \approx S_{iy}$ and ${\cal H}_{\tau}$ corresponds to an effective field which is {\it strictly\/} transverse. The appearance of such a term linear in $(b^{}_i - b^{\dagger}_{i}) $ signals that there is no time independent solution.  

Comparing ${\cal H}_{\tau}$ with the definition of $\hat p_z$ makes evident that ${\cal H}_{\tau} \propto j_s \hat p_z$, i.e., the current {\it only\/} couples to the collective coordinate of the wall via its momentum. The effect of ${\cal H}_{\tau}$ is then  to put the wall in motion along the $z$-direction, and it is necessary to study the time dependent Schr\"odinger equation: $i\hbar (\partial /\partial t) \psi(\vec r_i, t) = {\cal H} \psi(\vec r_i, t)$. The effect of adding the rotations, $r$, which displace the wall is $i\hbar (\partial /\partial t)  \to i\hbar r (\partial /\partial t) r^{-1} = i\hbar (\partial /\partial t) - \hbar( \partial \theta_i / \partial t )M_{iy}$, where $\vec M_i = \vec S_i + \vec s_i$ is the {\it total\/} spin  angular momentum. This  generates  a second  purely transverse field term in the effective Hamiltonian:
$$
\hbar \sum_i \frac{\partial \theta_i }{ \partial t}  M_{iy} 
= \hbar \frac{M}{S}\sum_i \frac{\partial \theta_i }{ \partial t} (2S)^{1/2}  (b^{}_i - b^{\dagger}_{i})
$$
using the fact that $(\vec M_i/M) = (\vec S_i/S)$, i.e., that all  magnetizations are parallel. Thus when 
\begin{equation}
\frac{\partial \theta_i }{ \partial t} = v_0 \frac{\partial \theta_i }{ \partial z};
\ \ \ \ v_0 = \frac{j_s a^3 }{2 M}.
\label{sept}
\end{equation}
the effective fields generated by the spatial and temporal rotations of the axes of quantization cancel each other. Given  $\theta_i=\theta(z_i) \equiv 2\cot^{-1} e^{-(z_i/w)}$ for $j_s=0$, the new {\it ground state\/} has $\theta_i=\theta(z_i - v_0t)$, i.e., the wall moves without distortion and without tilting or twisting. 
It is easy to show that the result $v_0 = (j_s v_c /2 M)$ reflects the conservation of the $z$-component of the total angular momentum, i.e., that the net spin current, $2j_s$ carried towards the wall by the electrons equals the change in the angular momentum, $j_S = Mv/v_c$, of the wall due to its motion. Since the conduction electrons are polarized $j_s = pj$ is related to the charge current $j$ by some material determined parameter $p$.

It is to be observed that by making these specific time dependent rotations the {\it net\/} transverse magnetic field has vanished from the problem. Thus the conditions for equilibrium in rotating frame are {\it identical\/} to those for the static problem, when $j_s=0$, and in particular  $\partial {\cal V}/\partial \phi_i=0$ since this is an equilibrium condition. In order to verify this point of divergence with Bazaliy et al.\cite{9}, it is useful to calculate this derivative directly. This requires the full SU(2) transformations. After some algebra the result is, Eqn.~(\ref{quatre}) with, 
$ \vec A= 
(1/2)  (\nabla \theta_i) {\sigma_y}^\prime 
+ 
(1/2) \sin \theta_i(\nabla \phi_i) {\sigma_x}^\prime
+ (1-\cos \theta_i)(\nabla \phi_i) \sigma_z^\prime
$
where the ${\vec \sigma}^\prime $ are defined in the local frame of reference. Directly, both new terms are  zero for a simple domain wall for which $\nabla \phi_i =0$, and  $\partial {\cal V}_{}/\partial \phi_i = 0$\cite{f3}. 
 
In terms of the Landau-Lifshitz equations, in the laboratory frame,
\begin{equation}
\frac{D \vec M }{D t} 
\equiv 
\frac{\partial \vec M }{ \partial t} - (\vec j_s \cdot \nabla) \vec M= g \mu_B \vec M \times \vec B
\label{huit}
\end{equation}
where the effective field is ($K_\perp=0$ for simplicity):
\begin{equation}
\vec B \equiv \frac{\partial {\cal V}}{\partial \vec M} = JS^2 a^2 \nabla^2\vec  M + \frac{2A}{M^2}(\hat z\cdot \vec M)\hat z .
\label{neuf}
\end{equation}
This is  formally similar  to the  result  obtained by Bazaliy et al.\cite{9}, in the half metal limit, {\it except\/} for  the fact that the term proportional to $j_s$ does {\it not\/} arise from a finite $\partial {\cal V}/\partial \phi$, via Eqn.~(\ref{neuf}). Also this torque transfer term arises from $\Delta t^\perp_{ij}$ which is simply {\it not\/} included in that work. Equation~(\ref{huit})  defines the  ``paticle derivative" $D\vec M/Dt$, i.e., that at a fixed position in the {\it moving\/} wall.

This different origin of the torque transfer term is of particular importance when relaxation is accounted for. With $j_s=0$, relaxation is traditionally included  through a Gilbert  term $-(\alpha/M)\vec M \times (\partial \vec M/\partial t)$, parameterized by $\alpha$, and which  assures  the system relaxes until $\vec M$ is parallel to the internal field $\vec B$, and which corresponds to an {\it absolute minimum\/} of ${\cal V}\{\theta_i,\phi_i\}$. This is  an evident requirement of  the {\it second law of thermodynamics}. In order that this law be satisfied when $j_s$ is finite, the Gilbert term must involve rather $D\vec M/Dt$ and the Landau-Lifshitz-Gilbert equations are,
\begin{equation}
\frac{D \vec M }{ D t} = g \mu_B \vec M \times \vec B -\frac{\alpha}{M}  \vec M \times \frac{D \vec M}{D t}.
\label{douze}
\end{equation}
with the present Eqn.~(\ref{neuf}).

The justification, at the microscopic level,  of this form of relaxation is complicated. For the micron sized system it is widely assumed that the principal loss mechanism corresponds spin-orbit scattering of the conduction electrons\cite{7}. Above a few tens of degrees, the Yafet-Elliot \cite{15}, spin-orbit scattering off phonons is almost certainly dominant. However, here there is little interest in obtaining a detailed expression for $(1/\tau)$ and so rather considered is the simplest model,
$$
{\cal H}_{\rm so} = i\lambda \sum_i \sum_{\vec k \vec k^\prime}\sum_{\sigma \sigma^\prime}
e^{i(\vec k-\vec k^\prime)\cdot \vec r_i}
\vec \sigma_{\sigma \sigma^\prime}\cdot (\vec k \times \vec k^\prime)c^\dagger_{\vec k \sigma}
c^{}_{\vec k^\prime \sigma^\prime}
$$
which represents such scattering off impurities, located are $\vec r_i$, and which enters the theory in the same manner as does phonon scattering. In fact, if the concentration $c$ of ``impurities" is taken to be $\sim  T/T_D$ where $T_D$ is the Deby temperature, this mimics the effect of phonons at modest temperatures. The problem is treated in the {\it rotating frame\/} since this reduces to zero the gauge fields created by the current $j_s$. Now in ${\cal H}_{\rm so}$ it is necessary to make the replacement  $\vec \sigma \to r \vec \sigma r^{-1}$.  The angular speed of rotation, $\sim \pi v_0/w \sim 10^3$rads/s, is by far the lowest frequency in the problem and hence it can be assumed that, $\omega_0 = \dot \theta_i$ is a constant during the relaxation process. Further, in practice, the wall width $w$ is large compared to the mean free path, and so the spatial dependence of $\omega_0$ can be ignored. The problem is then very similar to the linear response to an applied radio frequency field. In the context of dilute magnetic alloys, this has been studied in some detail\cite{16}. The Gilbert term is found to be of the correct form with, in the present notation, $\alpha = 1/(\omega_s \tau)$ where $\omega_s = g\mu_B B$ and,
$
(1/\tau)= W_{\uparrow \to \downarrow} - W_{\downarrow \to \uparrow} ,
$
and where, e.g., $W_{\uparrow \to \downarrow} $ is the transition rate for conduction electrons to flip from up to down spin given essentially by the ``Golden Rule". Assuming a  random  concentration of ``impurities", e.g., $W_{\uparrow \to \downarrow} = c(\pi \lambda^2/\hbar) \sum_{\vec k \vec k^\prime}[(k_z k_x^\prime - k_x k_z^\prime)^2\delta(\epsilon_{\vec k} - \epsilon_{\vec k^\prime}) +(1/4)( (k_y k_z^\prime - k_z k_y^\prime)^2 + (k_x k_y^\prime - k_y k_x^\prime)^2) \delta(\epsilon_{\vec k} - \epsilon_{\vec k^\prime}-\hbar \omega_0 ))
+(1/4)( (k_y k_z^\prime - k_z k_y^\prime)^2 + (k_x k_y^\prime - k_y k_x^\prime)^2) \delta(\epsilon_{\vec k} - \epsilon_{\vec k^\prime}+\hbar \omega_0 ))]$, where $\epsilon_{\vec k}$ is the energy of an electron with wave vector $\vec k$. The only effect occasioned by the passage to  the rotating frame is the appearance of $\hbar \omega_0$  in the delta functions with quantum corrections to  the form of the Gilbert term which are negligible since $\hbar \omega_0
\ll k_B T$. The modifications to the expression for $(1/\tau)$ are also negligible since, even for the true phononic mechanism,  $\hbar \omega_0$ is much smaller than the relevant  energy scale  $k_B T$.

This work was supported by a Grant-in-Aid for Scientific Research on
Priority Areas from the Ministry of Education, Science, Culture and
Technology of Japan, CREST and NAREGI. SEB was on a temporary leave from the
Physics Department, University of Miami, FL, U.S.A. and wishes to
thank the members of IMR for their kind hospitality.


\begin{thebibliography}{16}
\bibitem{f1} When $K_\perp =0$ there is a second rotational symmetry about the $z$-axis with $S_{iz}$, in the laboratory frame, as the local U(1) generator. It turns out that the single $b^\dagger_w$ is also the Goldstone boson for this symmetry.
\bibitem{f2} One of the interests in the spintronics field is the possibility of exerting forces on, and pinning of, a domain wall. The presence of $x^\prime$ in the expression for $J_s$ indicates that this potentially can be modified and modulated dynamically in, .e.g., a field effect device.
\bibitem{f3} However $(1-\cos \theta_i)(\nabla \phi_i) $ is a Berry phase which instead might be written as $\phi_i \sin \theta_i (\nabla \theta_i) $, implying an energy  $\sim j_s \phi_i$ and a finite $\partial {\cal V}_{}/\partial \phi_i$. These two fashions in which to write this phase {\it should\/} be related by a gauge transformation, i.e., $\vec A \to \vec A - \nabla f(\theta_i,\phi_i)$ where $f = (1-\cos \theta_i)\phi_i{\sigma_{z}}^\prime$. This connection however fails since there is a large surface term at the end with  $\theta_i = \pi$. The SU(2) rotations $r(\theta_i, \phi_i) \equiv e^{i\phi_i s_{z}} e^{i\theta_i s_{y}} e^{-i\phi_i s_{z}} $ used here are {\it special\/} in that the gauge fields $\vec A$ are zero far from the wall it creates and such surface terms are absent.
\end{thebibliography}
\end{document}